\documentclass[showpacs,preprintnumbers,prl,amsmath,amssymb,superscriptaddress, twocolumn]{revtex4}
\usepackage{dcolumn}
\usepackage{bm}
\usepackage{epsfig}

\begin{document}

\title{Enhancement of the Superconducting Transition Temperature under Pressure in \\Rare-Earth Doped Ca$_{1-x}$La$_x$Fe$_2$As$_2$ ($x=0.27$)}

\author{Swee~K.~Goh}
\author{Lina~E.~Klintberg}
\author{J.~M.~Silver}
\author{F.~M.~Grosche}
\affiliation{Cavendish Laboratory, University of Cambridge, J.J. Thomson Avenue, Cambridge CB3 0HE, United Kingdom}

\author{S.~R.~Saha}
\author{T.~Drye}
\author{J.~Paglione}
\affiliation{Center for Nanophysics and Advanced Materials, Department of Physics, University of Maryland, College Park, MD 20742 USA}

\author{Mike~Sutherland}
\affiliation{Cavendish Laboratory, University of Cambridge, J.J. Thomson Avenue, Cambridge CB3 0HE, United Kingdom}

\date{July 4, 2011}

\begin{abstract}
We report measurements of the pressure dependence of the superconducting transition temperature $T_c$ in single crystal samples of the rare-earth doped superconductor Ca$_{0.73}$La$_{0.27}$Fe$_2$As$_2$. We track $T_c$ with two techniques, via in-plane resistivity measurements and with a resonant tunnel diode oscillator circuit which is sensitive to the skin depth. We show that initially $T_c$ rises steeply with pressure, forming a
superconducting dome with a maximum $T_c$ of $\sim 44~\rm K$ at $20~{\rm kbar}$. We discuss this observation in the context of other electron-doped iron pnictide superconductors, and conclude that the application of pressure offers an independent way to tune $T_c$ in this system.

 \end{abstract}


\maketitle

The superconducting properties of the FeAs-based superconductors have been shown to be closely correlated with crystal structure. The dependence of $T_c$ on pnictogen bonding for instance was established early on \cite{Lee08,Kimber09}, reaching a maximum when the As-Fe-As bond angles are close to the ideal value of 109.47$^\circ$ in many materials \cite{Johnston10}. In the antiferromagnetic `parent compounds', the onset of spin density wave magnetic order is accompanied by a tetragonal to orthorhombic structural phase transition \cite{Zhao08,Nandi10}, which competes with superconductivity. Tuning of the structural parameters offers insight into the role that both magnetism and crystal structure play in stabilizing superconductivity.  This is achieved by several routes - the application of hydrostatic pressure, the substitution of isovalent atoms to create chemical pressure, or by the substitution of aliovalent atoms which can also change the doping level \cite{Paglione10}. 

The antiferromagnetically ordered AFe$_2$As$_2$ (A=Ca,Ba,Sr) materials are particularly interesting in this regard. Possessing a ThCr$_2$Si$_2$ crystal structure, these `122' compounds may be tuned to access a rich variety of phenomena. In CaFe$_2$As$_2$ neutron scattering and x-ray studies under hydrostatic pressure have shown that $T_N$ is rapidly suppressed, and at $p =$ 3.5 kbar the system undergoes a dramatic structural phase transition into a non-magnetic `collapsed' tetragonal (cT) phase \cite{Kreyssig08, Goldman09}, with a greatly reduced unit cell volume. Near this pressure, superconductivity emerges with a maximum $T_c$ of  $\sim$ 12 K \cite{Torikachvili08, Alireza09}, similar to the phase diagram recently reported for BaFe$_2$As$_2$ \cite{Mittal11}. In both cases, it is expected that uniaxial components of stress may play a role in determining $T_c$ \cite{Lee09,Torikachvili09,Baek09,Yu09,Duncan10}

An alternate route to superconductivity in the AFe$_2$As$_2$ system is through chemical substitution. A suppression of antiferromagnetic order is seen with both hole doping in Ba$_{1-x}$K$_x$Fe$_2$As$_2$ \cite{Rotter08,Chen09} and with electron doping in BaFe$_{2-x}$Co$_{x}$As$_2$  \cite{Sefat08,Chu09}, resulting in a superconducting state with a high $T_c$, reaching 38 K in the case of Ba$_{0.6}$K$_{0.4}$Fe$_2$As$_2$.  In both cases there is a small region of coexistence between the magnetic and superconducting phases, and the collapsed tetragonal state is not realized.

 The similarities (and differences) of the doping phase diagrams with those seen in the pressure studies has raised the question of whether the mechanism for enhancement of $T_c$ is the same in both cases \cite{Drotziger10, Klintberg10, Ahilan09}. In other words, is it the change in lattice structure that controls the appearance of superconductivity, or does the change in charge doping play a dominant role? 

Very recently a new approach to doping in the AFe$_2$As$_2$ system was reported  \cite{Saha11}, with trivalent rare earth elements La, Ce, Nd and Pr substituted for divalent Ca in CaFe$_2$As$_2$. This was subsequently confirmed by other groups \cite{Qi11,Lv11}. The substitution suppresses magnetic order and results in a collapsed tetragonal phase at ambient pressures in the case of Nd and Pr. Surprisingly, rare earth substitution also produces superconductivity at temperatures as high as 45 K, the highest yet reported for the AFe$_2$As$_2$ system. This $T_c$ is more than four times higher than the maximum $T_c$ achieved under non-hydrostatic pressure in CaFe$_2$As$_2$ \cite{Torikachvili08}. 

In this report we investigate whether $T_c$ may be increased further in the Ca$_{1-x}$R$_{x}$Fe$_2$As$_2$ (R=rare earth) system by applying pressure to the La member of the series, which carries no 4$f$ moment. We report the pressure dependence of $T_c$ in Ca$_{1-x}$La$_{x}$Fe$_2$As$_2$ with $x=0.27$, which is overdoped in the sense that the La concentration is higher than that required for
optimum $T_c$ within this substitution series, namely $x=0.2$. At ambient pressure,  our sample with $x=0.27$ gives $T_c$ $\simeq$ 31 K \cite{Saha11}.

\begin{figure} \centering
              \resizebox{\columnwidth}{!}{
              \includegraphics[angle=0]{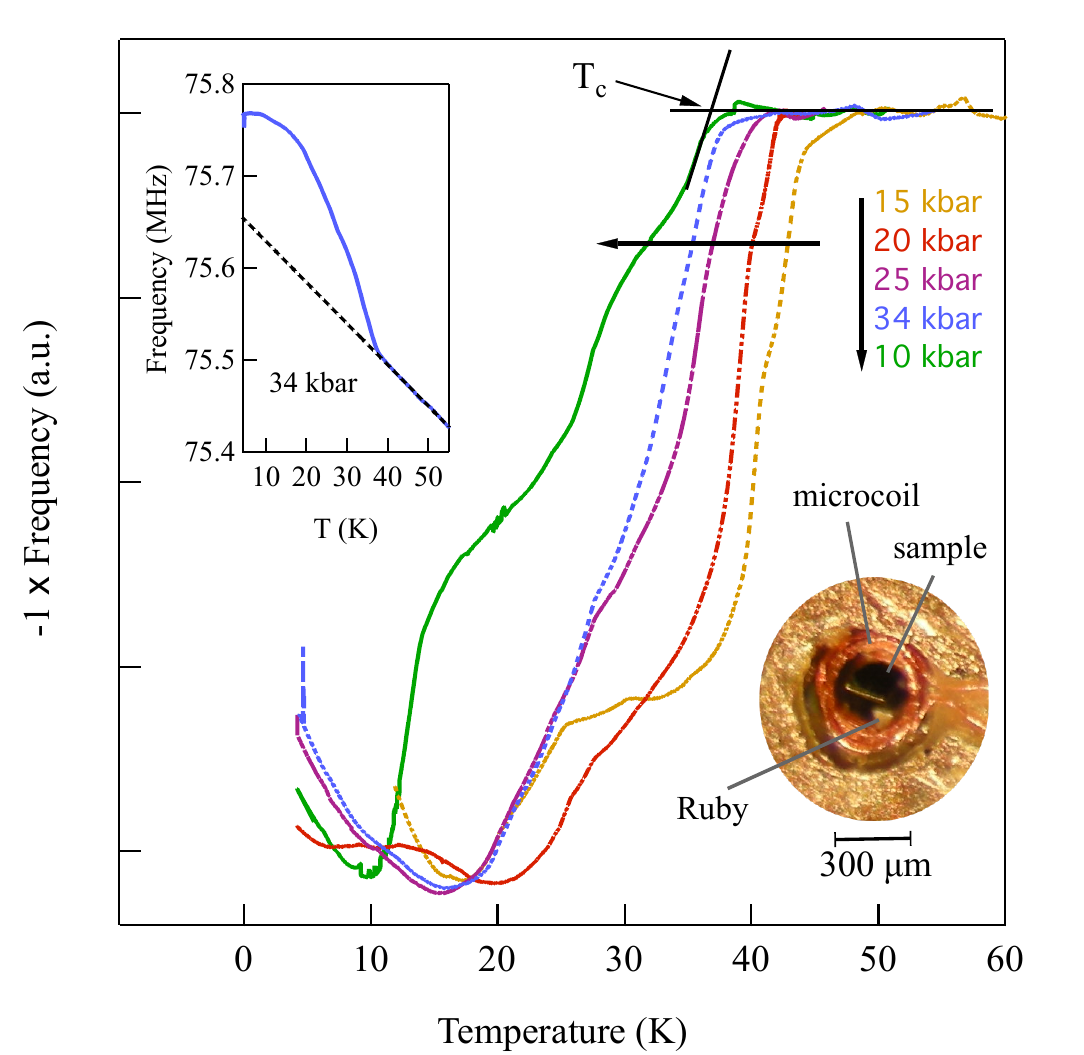}}
              \caption{\label{fig:1} (Color online) Temperature dependence of the TDO frequency at various pressures. Note that the vertical axis has been inverted. The upper inset shows the raw data and the estimation of the background. The lower inset is a photograph of a 10-turn microcoil with a diameter of 300~$\mu$m enclosing the sample and a ruby chip.}
       \end{figure}

Our samples were grown using a self flux technique \cite{Saha09}, yielding large single crystal samples of dimensions $\sim$ 10 $\times$ 10 $\times$ 0.1 mm$^3$. The La concentration of $x=0.27$ was determined using wavelength-dispersive x-ray spectroscopy and single-crystal x-ray diffraction measurements \cite{Saha11}. To detect the presence of superconductivity we use two techniques. The first is a conventional measurement of the four-wire electrical resistance of the sample, using a miniature piston-cylinder cell in a Quantum Design PPMS-9. The superconducting transition of lead was used as the manometer. The second technique tracks the resonant frequency of an oscillator formed by a tunnel diode (a BD-4 equivalent from MPulse) and a microcoil in the gasket hole of a Moissanite anvil cell (see the lower inset to Figure \ref{fig:1}). Ruby fluorescence spectroscopy was used for pressure determination. For both techniques, glycerin was used as the pressure transmitting fluid.


Figure \ref{fig:1} shows the temperature dependence of the  tunnel-diode oscillator (TDO) frequency with background removed at various pressures. The background is determined by a linear extrapolation of the data above the superconducting transition down to the lowest temperature of a particular run, as depicted in the upper inset to Figure \ref{fig:1}. Note that the vertical axis of the main panel is inverted. Therefore, similar to a recent high frequency study utilizing the microcoil setup \cite{Goh10}, the resonant frequency of the TDO \emph{increases} when the sample enters the superconducting state, and this sudden increase in the resonant frequency marks the onset of superconductivity. We define $T_c$ by the intersection of two lines. One, a linear fit to the data above the transition, and another a linear fit to the data just below the transition, as depicted in the figure.

       \begin{figure} \centering
              \resizebox{\columnwidth}{!}{
              \includegraphics[angle=0]{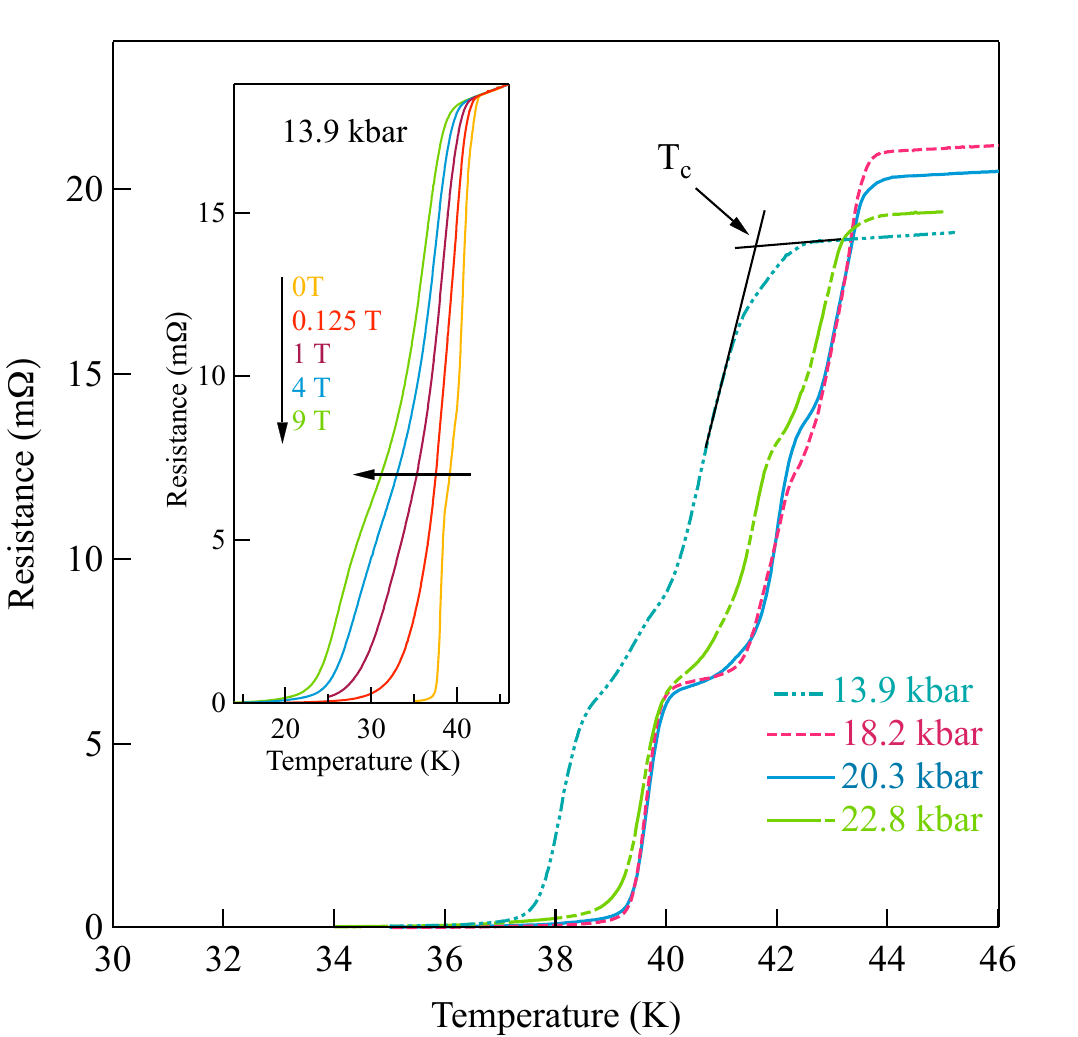}}
              \caption{\label{fig:2} Temperature dependence of resistance measured using a piston-cylinder cell. The inset shows the measurements of resistance with magnetic field applied along the $ab$-plane of the sample.}
        \end{figure}      
Measurements on a thin platelet of the in-plane resistivity indicate complete superconducting transitions (Figure \ref{fig:2}). Since all measurements were performed using the same pressure cell and the same sample, the dataset offers a meaningful comparison of the resistance just above $T_c$, $R(T\rightarrow T_c^+)$, at different pressures. Our data suggest that $R(T\rightarrow T_c^+)$ has a distinct maximum near the
pressure for maximum $T_c$.

When a magnetic field is applied, the superconducting transition becomes broader and $T_c$ decreases relatively fast at low field, but the reduction in $T_c$ slows down at high field (inset of Figure \ref{fig:2}). In other words $T_c$ is nonlinear, changing from a small slope at low field to a large slope at high field.
       \begin{figure} [!t]\centering
              \resizebox{\columnwidth}{!}{
              \includegraphics[angle=0]{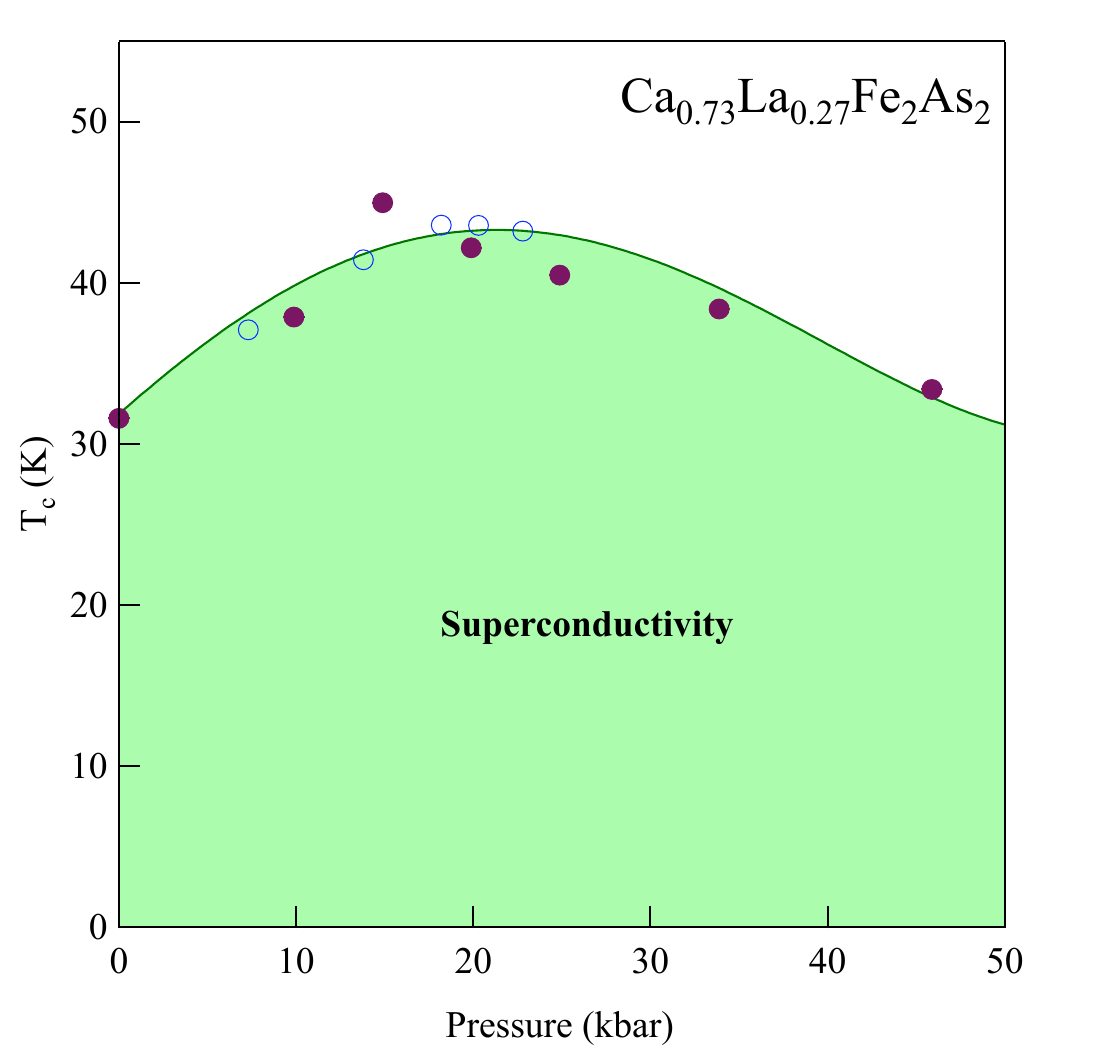}}
              \caption{\label{fig:3} Phase diagram showing the pressure dependence of $T_c$ constructed from resistance ({\Large{$\circ$}}) and TDO ({\Large{$\bullet$}}) measurements. The solid line is a guide to the eye.}
      \end{figure}
Using the data from both techniques, the pressure phase diagram can be constructed. We find that the superconductivity is very sensitive to pressure: $T_c$ rises from 31~K at ambient pressure, reaching a maximum value of $\sim$~44~K at about 20~kbar. This gives an average slope of  $\sim$~0.65~K/kbar, a value that is considerably higher than that in other electron doped 122 pnictides. 
In overdoped BaFe$_{1.8}$Co$_{0.2}$As$_2$ for instance,  $T_c$ was observed to increase with an initial slope of 0.065 K/kbar, then level off at higher pressures, growing a total of only 1~K with the application of 25 kbar \cite{Ahilan09}. In the underdoped regime, the pressure coefficient does become sizable, reaching 0.4 K/kbar for a sample with composition BaFe$_{1.92}$Co$_{0.08}$As$_2$ \cite{Ahilan09}.

Why does an overdoped sample of Ca$_{1-x}$La$_x$Fe$_2$As$_2$ display the pressure sensitivity characteristic of an underdoped sample of BaFe$_{2-x}$Co$_{x}$As$_2$? From a structural perspective, the replacement of Ca (atomic radius 126 pm) with La (130 pm) expands the $a$-axis lattice parameter, from 3.895 \AA\ for pure CaFe$_2$As$_2$ to $\sim$ 3.92 \AA\ for x=0.27 as measured at 250~K. The $c$-axis remains relatively unchanged, within the margins of error in the measurement \cite{Saha11}. Applying hydrostatic pressure to CaFe$_2$As$_2$ has a somewhat different effect, causing a significant shortening of the $c$-axis, from 11.75 \AA\ to 11.50 \AA\ by 10 kbar in the high temperature tetragonal phase (T') at room temperature. Over the same pressure range, the $a$-axis lattice constant  increases slightly, by 0.1\%\ \cite{Canfield09}.

It is therefore highly unlikely that the application of hydrostatic or nearly hydrostatic pressure tunes the Ca$_{1-x}$La$_{x}$Fe$_2$As$_2$ system in a way that simply mimics the structural changes that accompany doping. Since the $a$-axis lattice constant changes the most with doping, it would seem plausible that applying uniaxial pressure along this direction would tune the system in a similar manner. However, it is quite interesting to note that the maximum value of  $T_c$ = 44~K seen in our pressure experiments on Ca$_{0.73}$La$_{0.27}$Fe$_2$As$_2$ is very close to that seen at ambient pressure at optimally doped (43~K) Ca$_{0.8}$La$_{0.2}$Fe$_2$As$_2$ in this system \cite{Saha11, Qi11}.

Alternatively, one might consider the role of density fluctuations in boosting $T_c$. For undoped CaFe$_2$As$_2$ at room temperature, a volume collapse into the cT state is known to occur when the interlayer As-As separation approaches 3.0 \AA\ \cite{Saha11}. At sufficiently high pressures one might expect our sample to change into the cT state as well, but whether this occurs as a violent first order transition, as in CaFe$_2$As$_2$, now needs to be investigated.
In the case of materials which possess ground state crystal structures that are close in energy, soft lattice modes may develop when the transition temperature is suppressed towards $T=0$. It has been shown that such a density instability can create an effective interaction, which is enhanced in quasi-2D materials \cite{Monthoux04}. Structural studies under high pressure in the Ca$_{1-x}$La$_{x}$Fe$_2$As$_2$ system would be useful in clarifying the phase diagram and determining whether there is any support for such a scenario.
 
The proximity to the volume collapse transition raises the possibility of domain formation. High pressure neutron scattering studies in CaFe$_2$As$_2$ \cite{Prokes10} have shown that the T' phase coexists with the cT phase when the sample is subject to non-hydrostatic pressure conditions. The relative population of these two structures can be tuned with pressure, with the T' phase supporting superconductivity, which is generally thought to be absent in the cT phase \cite{Yu09}. This scenario leads to superconductivity with a reduced volume fraction, consistent with that reported so far in this system \cite{Saha11,Qi11,Lv11}. However it is still an open question as to why $T_c$ is so much higher in the present case than in the undoped system.

In summary, we have studied the pressure dependence of a sample of overdoped Ca$_{1-x}$La$_x$Fe$_2$As$_2$, and found a rapid rise in $T_c$ where one might naively expect a suppression. We conclude that the application of pressure in this system does not simply follow the expectation from doping, and offers an additional route to tune $T_c$ in this family of materials.


\begin{acknowledgments}

We would like to thank G. Lonzarich and P. Alireza for useful discussions. M.S. acknowledges 
support from the Royal Society, and S.K.G. from Trinity College Cambridge. Research at the University of Maryland was supported by the AFOSR-MURI (FA9550-09-1-0603) and the NSF (DMR-0952716)

\end{acknowledgments}


\end{document}